\documentstyle[12pt]{article}

\begin{document}

\title{Geometric models of (d+1)-dimensional relativistic rotating oscillators}


\author{Ion I. Cot\u aescu\\ 
       {\small \it The West University of Timi\c soara,}\\
       {\small \it V. P\^ arvan Ave. 4, RO-1900 Timi\c soara, Romania}}

\date{\today}

\maketitle

\begin{abstract}
Geometric models of quantum relativistic rotating oscillators in arbitrary 
dimensions are defined on backgrounds with deformed anti-de Sitter metrics. 
It is shown that these models are analytically solvable, deriving the formulas 
of the energy levels and  corresponding normalized energy eigenfunctions.   
An important property is that all these models have the same nonrelativistic 
limit, namely the usual harmonic oscillator.   

Pacs 04.62.+v
\end{abstract}
\


In general relativity the anti-de Sitter (AdS) spacetime is one of the most 
important and interesting pieces  since the AdS/Conformal field 
theory-correspondence \cite{M} has been discovered. It is known that, because 
of the high symmetry of AdS, the free motion of the scalar classical or 
quantum particles on this background has special features. There is a local 
chart  with a metric \cite{B} able to reproduce  the classical motion of an 
isotropic nonrelativistic harmonic oscillator (NRHO).  In other respects, 
the  energy levels of the quantum modes given by the Klein-Gordon equation 
are equidistant   \cite{I}. Thus the geometric models of free test 
classical or quantum particles  on  AdS backgrounds represent the relativistic 
correspondent of the NRHO. 

Two years ago, we have generalized this ideal model of relativistic oscillator 
(RO) to new families of models of RO in (1+1) and (3+1) dimensions 
\cite{C1,C2}. In general, the metrics of these models are deformations of some 
AdS or de Sitter metrics that produce oscillations and a specific relativistic 
rotation effect in the case of the (3+1)-dimensional models \cite{C2}. 
However, what is important here is that all these models have as  
nonrelativistic limit just the NRHO. 
We have studied in details the (1+1) \cite{C1} models finding that there are 
two kind of quantum RO with different properties, namely P\" oschl-Teller-like 
models (with deformed AdS metrics) and respectively Rosen-Morse-like models 
(with deformed de Sitter metrics) \cite{C3}. Moreover, we have shown that all 
the (1+1)-dimensional P\" oschl-Teller-like RO have similar  properties and the 
same $so(1,2)$ dynamical algebra \cite{C4}. The next step might be the study  of 
the algebras of the (3+1) RO for which we have obtained only the energy spectra 
and the wave functions up to normalization factors \cite{C2}. Fortunately, the 
space dimension of the model is not determinant for solving the Klein-Gordon 
equation. This means that these models can be analytically solved in any 
dimensions like the AdS one \cite{C5}. Thus we have the opportunity to 
completely solve the problem of the quantum modes of rotating RO in arbitrary 
dimensions and then turn to the study of their dynamical algebras.

Our aim is to present here only the method of solving the 
P\" oschl-Teller-like models in (d+1) dimensions. Our main objective is to 
derive the formula of the energy levels and to find the normalized energy 
eigenfunctions in spherical coordinates (and natural units with $\hbar=c=1$). 
Moreover, we show  that, like in (3+1) dimensions, the rotation of these 
(d+1) RO is a pure relativistic effect that vanishes in the nonrelativistic 
limit when all these models lead to the (d+1)-dimensional NRHO.

>From the theory of the (3+1) RO we understand that the background of any 
(d+1) RO  must be static and spherically symmetric (central) \cite{C1}. 
Therefore, the backgrounds of our (d+1) RO must have central  static 
charts with generalized spherical coordinates, $r, \theta_{1},...,
\theta_{d-1}$,  commonly related with the Cartesian ones ${\bf x}\equiv (x^{1}, 
x^{2},...,x^{d})$ \cite{T}. Here it is convenient to chose the radial 
coordinate such that $g_{rr}=-g_{00}$ since then the radial scalar product is 
simpler \cite{C4}. Starting with these options, we  define the metrics of our 
(d+1)-dimensional  P\" oschl-Teller-like RO as one-parameter {\em deformations} 
of the AdS metric given by the line elements 
\begin{equation}
ds^{2}=\left( 1+\frac{1}{\epsilon^2}\tan^{2}\hat\omega r\right) (dt^{2}-dr^{2}) 
-\frac{1}{\hat\omega^2}\tan^{2}\hat\omega r\, d\theta^{2}
\end{equation}
where we denote $\hat\omega=\epsilon\,\omega$, $\epsilon\in [0, \infty)$, and  
\begin{equation}
d\theta^{2}={d\theta_{1}}^{2}+\sin^{2}\theta_{1}{d\theta_{2}}^{2}\, ...\, 
+\sin^{2}\theta_{1}\,\sin^{2}\theta_{2}...
\sin^{2}\theta_{d-2}{d\theta_{d-1}}^{2} 
\end{equation}
is the usual line element on the sphere $S^{d-1}$. The deformation parameter 
$\epsilon$ determines the geometry of the background while $\omega$ remains 
fixed. It is clear  that for $\epsilon=1$ we obtain just the AdS metric 
(with the hyperboloid radius $R=1/\omega$) \cite{C5}. An interesting case is 
that of  $\epsilon\to 0$ when the line element, 
\begin{equation}
ds^{2}=(1+\omega^{2} r^{2}) (dt^{2}-dr^{2}) -
 r^{2}\, d\theta^{2}\,,
\end{equation}
defines a background where the relativistic quantum motion is similar to that 
of NRHO. This model will be called the {\em normal} RO. In general, the radial 
domain of any RO is  $D_{r}=[0,\, \pi/2\hat\omega)$ which means 
that the whole space domain is $D=D_{r}\times S^{d-1}$. 
For the models with $\epsilon\not=0$ the time  might satisfy the condition  
$t\in [-\pi/\hat\omega,\pi/\hat\omega)$ as in the AdS case but here we  
consider that $t\in (-\infty,\infty)$ which corresponds to the universal 
covering spacetimes of the original backgrounds.

In our models the oscillating test particle is described by a scalar quantum 
field $\phi$ of mass $M$, minimally coupled with the gravitational 
field. Its quantum modes are given by the particular solutions of the 
Klein-Gordon equation 
\begin{equation}\label{(kg)}
\frac{1}{\sqrt{g}}\partial_{\mu}\left(\sqrt{g}\,g^{\mu\nu}\partial_{\nu}\phi
\right) + M^{2}\phi=0\,, \quad
g=|\det(g_{\mu\nu})|\,. 
\end{equation} 
These  may be either square integrable functions or distributions on  $D$. 
In  both cases they must be orhonormal (in usual or generalized sense) with 
respect to the relativistic scalar product \cite{BD} 
\begin{equation}\label{(sp)}
\left<\phi,\phi'\right>=i\int_{D}d^{d}x\,\sqrt{g}\,g^{00}\,
{\phi}^{*}\stackrel{\leftrightarrow}{\partial_{0}} \phi'\,.
\end{equation}

The spherical variables of Eq.(\ref{(kg)}) can be separated by using  
generalized spherical harmonics, $Y^{d-1}_{l\,(\lambda)}({\bf x}/r)$. 
These are normalized eigenfunctions of the angular Lalpace operator \cite{T},  
\begin{equation}
-\Delta_{S}
Y^{d-1}_{l\,(\lambda)}({\bf x}/r) 
=l(l+d-2)\,Y^{d-1}_{l\,(\lambda)}({\bf x}/r)\,, 
\end{equation} 
corresponding to eigenvalues depending on the {\em angular} quantum number 
$l$ which  takes  the  values $0,1,2,....$ \cite{T}. The notation $(\lambda)$ 
stands for a collection of quantum numbers giving the multiplicity of these 
eigenvalues \cite{T}, 
\begin{equation}\label{(gamal)}
\gamma_{l}= (2l+d-2)\frac{(l+d-3)!}{l!\,(d-2)!}\,.
\end{equation}
We start with  (positive frequency) particular solutions of energy $E$,   
\begin{equation}\label{(udex)}
\phi^{(+)}_{E,l(\lambda)}(t,{\bf x})\sim (\cot\hat\omega r )^{\frac{d-1}{2}}\, 
R_{E,l}(r)\,Y^{d-1}_{l\,(\lambda)}({\bf x}/r)\, e^{-iEt}\,. 
\end{equation}
Then, after a few manipulation, we find  the radial equation   
\begin{equation}\label{(radeq)}
\left[-\frac{1}{\hat\omega^2}\frac{d^2}{dr^2}+\frac{2s(2s-1)}{\sin^{2}\hat\omega r}+
\frac{2p(2p-1)}{\cos^{2}\hat\omega r}\right]R_{E,l}=\nu^{2} R_{E,l}
\end{equation}
where 
\begin{equation}\label{(eqps)}
2s(2s-1)=\left(l+\frac{d}{2}-1\right)^{2}-\frac{1}{4}\,, \quad 
2p(2p-1)=\frac{M^2}{\hat\omega^{2}\epsilon^{2}}+\frac{d^{2}-1}{4}\,,
\end{equation}
and
\begin{equation}
\nu^{2}=\frac{E^2}{\hat\omega^2}-\left(1-\frac{1}{\epsilon^2}\right)
\left[\frac{M^2}{\hat\omega^2}-l(l+d-2)\right]\,.
\end{equation}
This equation gives radial functions,
\begin{equation}\label{(gsol)}
R_{E,l}(r)\sim \sin^{2s}\hat\omega r\cos^{2p}\hat\omega r
F\left(s+p-\frac{\nu}{2},
s+p+\frac{\nu}{2}, 2s+\frac{1}{2}, \sin^{2}\hat\omega r\right)\,,
\end{equation}
expressed in terms of Gauss hypergeometric functions \cite{AS} depending 
on the real parameters $s$,  $p$ and $\nu$.
The radial functions (\ref{(gsol)}) have good physical meaning only when 
the functions $F$ 
are polynomials selected by a suitable quantization condition since otherwise 
these are strongly divergent for $r\to \pi/2\hat\omega$. Therefore, 
we introduce the radial quantum number $n_{r}$ \cite{BL} and impose  
the quantization condition
\begin{equation}\label{(quant)}
\nu=2 (n_{r}+s+p)\,,\quad n_{r}=0,1,2,...
\end{equation}
In addition, we choose the boundary conditions of the {\em regular} modes 
given by the  positive solutions of Eqs.(\ref{(eqps)}),  
i.e. $2s=l+(d-1)/2$ and $2p=k-(d-1)/2$, where  we denote  
\begin{equation}
k=\sqrt{\frac{M^{2}}{\hat\omega^{2}\epsilon^2}+\frac{d^2}{4}}+ \frac{d}{2}\,.  
\end{equation}
This new parameter which concentrates all the other ones can be used as  
the main parameter of RO, like in the case of the (1+1) models where $k$ was 
just the weight of the irreducible representations of the $so(1,2)$ dynamical 
algebra \cite{C4}.

The last step is to define the {\em main} quantum number, $n=2n_{r}+l$, which 
takes the  values, $0,1,2,...$,  giving the energy levels 
\begin{equation}\label{(e)}
E_{n,l}^{2}=\hat\omega^{2}(k+n)^{2}+\hat\omega^{2}(\epsilon^{2}-1)
\left[k(k-d)-\frac{1}{\epsilon^2} l(l+d-2)\right].
\end{equation}
If $n$ is even then $l=0,2,4,...,n$ while for odd $n$ we have $l=1,3,5,...,n$. 
In both cases the degree of degeneracy of the level $E_{n,l}$ is given by 
(\ref{(gamal)}). The last term of (\ref{(e)}) is just the rotator-like term 
that gives the behavior of rotating oscillator. Obviously, this does not 
contribute in the AdS case when $\epsilon=1$. On the other hand, we observe 
that the rotation effect vanishes in the nonrelativistic limit since 
the rotator-like term decreases as $1/c^2$ when $c\to \infty$ (in usual units).   
Now it remains only to  express (\ref{(gsol)}) in terms of Jacobi 
polynomials and to normalize them to unity with respect to (\ref{(sp)}). 
The final result is  
\begin{eqnarray}
\phi^{(+)}_{n,l(\lambda)}(t,{\bf x})&=&N_{n,l}
\sin^{l}\hat\omega r\cos^{k}\hat\omega r\\ 
&&\times {P_{n_{r}}^{(l+\frac{d}{2}-1,\,k-\frac{d}{2})}}(\cos 2\hat\omega r)
\,Y^{d-1}_{l(\lambda)}({\bf x}/r)\,e^{-iE_{n,l}t},\nonumber 
\end{eqnarray}
where  
\begin{equation}
N_{n,l}=\left[\frac{\hat\omega^d}{E_{n,l}}\frac{n_{r}!\,
(2n_{r}+k+l)\Gamma(n_{r}+k+l)}
{\Gamma (n_{r}+l+\frac{d}{2})\Gamma(n_{r}+k+1-\frac{d}{2})}\right]
^{\frac{1}{2}}\,.
\end{equation}
Particularly, for $\epsilon=1$ we recover the result we have recently obtained 
in AdS case \cite{C5}.

According to the above result,  we can say that in the models with 
$\epsilon\not=0$ the particles have the same space behavior. However, the 
situation is different for  $\epsilon\to 0$.
It is not difficult to show that in this limit, when $k$ increases as  
$M/\omega\epsilon^{2}$, we obtain the energy levels of the normal RO   
\begin{equation}\label{(en)}
\lim_{\epsilon\to 0} E^{2}_{n,l}={\stackrel{o}{E\,}}^{2}_{n,l}=
M^{2}+2\omega M\left( n+\frac{d}{2}\right)+\omega^{2}\,l(l+d-2)
\end{equation}
which have their specific rotator-like terms (of the order $1/c^2$). The 
corresponding energy eigenfunctions can be expressed in terms of Laguerre 
polynomials as,
\begin{eqnarray}
\lim_{\epsilon\to 0}\phi^{(+)}_{n,l(\lambda)}(t,{\bf x})&=&
\stackrel{o}{\phi\,}^{(+)}_{n,l(\lambda)}(t,{\bf x})=
\left[\frac{(\omega M)^{l+\frac{d}{2}}}{\stackrel{o}{E}_{n,l}}
\frac{n_{r}!}{\Gamma(n_{r}+l+\frac{d}{2})}\right]^{\frac{1}{2}}\label{(phio)}\\
&&\times  r^{l}e^{-\omega Mr^{2}/2}\,
L_{n_{r}}^{l+\frac{d}{2}-1}(\omega M\,r^{2})
Y_{l(\lambda)}^{d-1}({\bf x}/r)e^{-i\stackrel{o}{E}_{n,l}\,t}\,.
\nonumber
\end{eqnarray}
It is remarkable that these  wave functions coincide to those of NRHO up to the 
factor $1/\sqrt{2\stackrel{o}{E}_{n,l}}$ which appears since the definition of 
the relativistic  scalar product is different from that of the nonrelativistic 
one. Moreover, one can verify that in the nonrelativistic limit, 
(for $c\to \infty$ and very small nonrelativistic energies $\tilde E=E-Mc^{2}$, 
in usual units), the energy levels (\ref{(en)})  become just those of the NRHO, 
i.e. $\tilde E_{n}=\omega (n+d/2)$.
On the other hand, we observe that the nonrelativistic limit of any RO can 
be calculated taking first $\epsilon\to 0$ and then $c\to \infty$. 
Therefore, we can conclude that  all the models of rotating RO studied here 
lead to the (d+1)-dimensional NRHO in the nonrelativistic limit.

\end{document}